# Present status of theoretical understanding of charge changing processes at low beam energies

D. K. Swami and T. Nandi

Inter University Accelerator Centre, JNU New Campus, New Delhi-110067, India.

A model for the evaluation of charge-state distributions of fast heavy ions in solid targets is being developed since late eighties in terms of ETACHA code. Time to time it is being updated to deal with more number of electrons and non-perturbative processes. The calculation approach of the recent one, which is formulated for handling the non-perturbative processes better, is different from the earlier ones. However, the experimental results for the projectiles up to 28 electrons can be compared with the predictions from any versions of ETACHA code. Though earlier versions are not meant for the non-perturbative cases, but the detail comparison suggests that predictions from an earlier version is somewhat superior to that of the recent version. However, certain difference up to 4 units of charge found between the earlier version and experimental results on the mean charge states and charge state distributions is attributed to nonradiative electron capture taking place at the exit surface in the influence of wake and dynamic screening effects. This can be a possible mechanism of multiply charge formation in the electrospray ionization of big molecules.

## 1. Introduction:

Charge changing processes of projectile ions traversing solid or gaseous targets has been a subject of interest for more than 70 years [1] for achieving better fundamental understanding and numerous practical applications [2]. The process is highly intricate due to various physical phenomena including ionization, excitation, radiative decay, Auger decay, electron decay, radiative and non-radiative electron capture, etc. Variation of charge state fractions (CSF) versus charge state called charge state distribution (CSD) is used in both experiments and theories to understand the charge changing processes in detail. Several extensive reviews can be found in the literature [[3][4][5][6][7]]. These reviews provide the theoretical background and experimental techniques as well as data collected until the date of the corresponding publications on the CSD. The CSDs are normally measured by the standard electromagnetic measurements [4] and in order to understand the measured mean equilibrium charge state data, many semi-empirical formulas such as Thomas-Fermi Model, Bohr Model, Betz Model, Nikolaev-Dmitriev Model, To-Drouin Model, Shima-Ishihara-Mikumo Model, Itoh Model, Ziegler-Biersack-Littmark Model, Schiwietz Model have been developed [8] in tune with the experimental results from electromagnetic measurements. However, the semi-empirical formulas fail to estimate the non-equilibrium charge states and equilibrium foil thickness. Hence, a dedicated effort has been put in developing a model [9].

The model calculation of projectile charge-state distributions as a function of penetration depth x in a solid target is performed by solving a set of differential equations to account for different cross sections responsible for corresponding atomic processes such as ionization, excitation, radiative and non-radiative electron capture. Subsequently, numerical calculations of the model have been put in a code in the name of ETACHA [10] and time to time they have extended the model more and more useful for the multi-electron systems and non-perturbative processes [11]. The first version is now called ETACHA23[10], which makes use of the plane-wave Born approximation (PWBA) for ionization and excitation, whereas the continuum distorted-wave (CDW) approximation for the capture cross sections. This code is limited to 28 projectile orbital states of electrons. The same theoretical approach is continued in the versions ETACHA3 and ETACHA34, however, they can take higher number of projectile orbital states of electrons. In next step, beside the number of states that needs to be included to handle projectile ion states, the extensions of the ETACHA code is proposed also to deal with collision systems in the nonperturbative regime. This development is required because the above theoretical approaches are supposed to fail in reproducing the experimental results. Hence, ETACHA4 has been developed with the theoretical description of the ionization and excitation processes by introducing the continuum distorted-wave-eikonal initial state (CDW-EIS) approximation [[12][13]] and the symmetric eikonal (SEIK) model [[14][15]], respectively. Note that in these two formulations, the multielectronic projectile is reduced to an effective monoelectronic one, assuming that the nonionized electrons remain frozen in their initial orbitals. Instead of this, the CDW-EIS illustrates better than the PWBA for the experimental results of total ionization cross sections of He as a function of projectile velocity induced by different collision partners, namely, $H^+$, $He^{2+}$, $Li^{3+}$ in the energy range of 5-3000 keV/u. Further, the SEIK explains $1s-2p$ excitation cross section for $Ar^{17+}$ ions at beam energy 13.6 MeV/u as a function of exciting target atomic number better than the PWBA. Further, nonradiative capture cross sections can be evaluated by either the CDW or the relativistic eikonal approximation (CEIK). Though these two calculations differ from each other in the low energy side (<30 keV/u), but no difference is found in the higher energy side. Nevertheless, the CEIK is computationally much simpler than the CDW. Hence, ETACHA4 utilizes CEIK in place of CDW.

Above discussions suggest that the measured CSD must be following better the ETACHA4 predictions for the collisions in the energy range 0.5≤E≤8 Mev/u, however we notice just the opposite. Though solving this riddle is a theoretical challenge, but it is out of scope of the present work. Rather we define the problem here in greater detail so as to the community gets aware of such an unusual fact. To do so we proceed in the following way. We show in section 2.1 that the empirical formulae fail to reproduce many observed facts satisfactorily. Hence, a model calculation through ETACHA code [[9][10][11]] has been used to compare the mean charge state ($\bar{q}$) of various ions through carbon foil. The ETACHA4 results have been discussed in terms of the projectile perturbation parameters. In section 2.2, we present the variation of $\bar{q}$ with beam energy for different targets. Equilibrium foil thickness has been discussed in Section 2.3. Section 2.4 deals with the shell effect. In section 2.5, a discussion of charge state fraction with ion beam energy is elaborated for projectiles $Z_p$ ≥ 17 with two examples through argon and copper ion beams. In section 2.6, we make a thorough test on the predicting power of ETACHA codes on the equilibrium target thickness of various elements at different beam energies, for example a light element beryllium to a heavy element gold. Effect of electron capture phenomenon and surface potential at the exit surface on $\bar{q}$ has been discussed in section 2.7. Finally, we summarize the entire result in a tabular form to conclude that ETACHA23 predictions are somewhat better than that of ETACHA4 and non-radiative capture processes is vital for a theory to reproduce the experimental results.

## 2. Results and discussions:

### 2.1 Effect of the projectile atomic number ($Z_p$) on the equilibrium $\bar{q}$

Fig. 1 shows a comparison between experimental data (taken from [6]) and theoretical results from ETACHA codes [[10],[11]] for equilibrium $\bar{q}$ against beam energy. Here the projectiles vary but the target remains fixed with the carbon foil. For first three projectile ions from carbon to fluorine, there is slight difference between the experimental and calculated data; ETACHA predictions are a bit over estimated. ETACHA23 continues with the overestimations. In contrast the agreement is fairly good with the ETACHA4 for the ions from silicon to sulfur. However, the scenario changes towards opposite side; with the heavier ions from chlorine to copper; ETACHA4 predictions are underestimated. It can be seen that the agreement with the copper is far worse than the other ions.

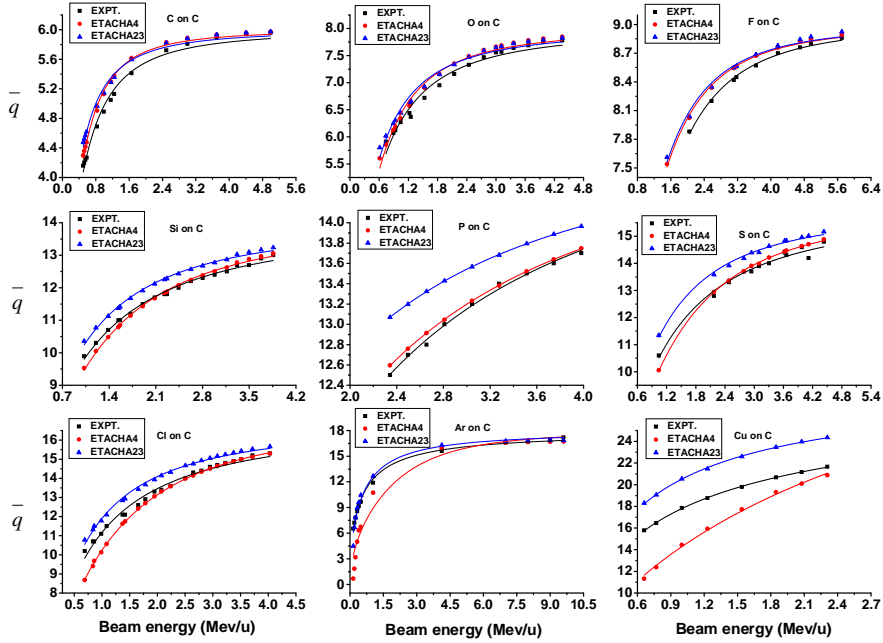

Fig. 1: The equilibrium $\bar{q}$ vs beam energy for different projectiles on the carbon target. Experimental data are taken from [6]. Both experimental and theoretical values are fitted with the equation (1).

The observed as well as theoretical trend of equilibrium $\bar{q}$ vs beam energy shown in Fig. 1 can be fitted with an empirical relation of $\bar{q}$ vs E similar to the equation given by Nikolaev and Dmitriev [16] as follows:

$$\bar{q} = Z_p \left(1 + X^{\frac{-1}{a}}\right)^{-a} \qquad (1)$$

$$X = b \frac{\sqrt{E/M_p}}{Z_p^{\,c}} \qquad (2)$$

where E is the kinetic energy of the projectile in MeV, and $Z_p$, $M_p$ are the atomic number and mass number of the projectile ions, respectively and a, b, and c are the fitting parameters. Nikolaev and Dmitriev [16] have taken the values of a, b, and c for carbon target 0.6, 3.86, and 0.45, respectively. However, we notice these parameter values do not fit all the collision systems. Accordingly, we have fitted each system with free parameters and the values so obtained are given in Table 1. It can be seen that parameter *a* in the experiments vary from 0.23 to 0.61, whereas it takes shorter range 0.23 to 0.48 in ETACHA23 and 0.18 to 0.31 in ETACHA4 predictions instead of 0.6 [16]; parameter b ranges from 0.76 to 3.89 in the experiments, 0.76 to 3.58 in ETACHA23 and 0.25 to 3.20 in ETACHA4 predictions instead of 3.86 [16]; parameter c ranges from 0.003 to 0.59 in the experiments, 0.004 to 0.53 in ETACHA23 and -0.33 to 0.57 in ETACHA4 predictions instead of 0.45 [16]. Variation of the fitting parameters from system to system limits the use of the empirical relations in predicting the equilibrium $\bar{q}$ for an unknown system and hence, it recommends for use of an ab initio theory [[9],[10],[11]].

Table1: The fitting parameters for an empirical relation (Eqn. 1) of equilibrium $\bar{q}$ vs E for different projectiles on carbon foil target.

| Proj. | Target | Beam Energy Range (MeV) | a | | | b | | | c | | |
|---|---|---|---|---|---|---|---|---|---|---|---|
| | | | Expt. | ETA23 | ETA4 | Expt. | ETA23 | ETA4 | Expt. | ETA23 | ETA4 |
| C | C | 6.0 - 59.9 | 0.32 | 0.32 | 0.27 | 1.43 | 1.61 | 1.51 | 0.16 | 0.15 | 0.18 |
| O | C | 11.8 - 69.8 | 0.33 | 0.33 | 0.30 | 3.28 | 3.15 | 3.08 | 0.59 | 0.53 | 0.57 |
| F | C | 38.9 - 108.3 | 0.23 | 0.23 | 0.23 | 0.97 | 0.84 | 0.80 | 0.12 | 0.03 | 0.01 |
| Si | C | 28.7 - 108.1 | 0.36 | 0.33 | 0.30 | 0.83 | 0.86 | 0.74 | 0.003 | 0.006 | 0.003 |
| P | C | 72.6 - 123.4 | 0.30 | 0.35 | 0.31 | 1.8 | 2.33 | 2.99 | 0.35 | 0.36 | 0.52 |
| S | C | 33.3 - 141.8 | 0.35 | 0.32 | 0.28 | 0.76 | 0.79 | 0.77 | 0.02 | 0.004 | 0.06 |
| Cl | C | 24.1 - 141.1 | 0.41 | 0.39 | 0.29 | 1.77 | 1.84 | 0.25 | 0.28 | 0.27 | -0.33 |
| Ar | C | 6.0 - 384.0 | 0.54 | 0.46 | 0.25 | 1.16 | 0.76 | 0.48 | 0.06 | -0.06 | -0.01 |
| Cu | C | 43.0 - 150.0 | 0.61 | 0.48 | 0.18 | 3.89 | 3.58 | 3.20 | 0.44 | 0.39 | 0.55 |

When a single theoretical approach attempts to tackle collision systems in the perturbative as well as nonperturbative regime, very often it fails to reproduce the experimental results. In order to have an idea whether collision system is within perturbative or nonperturbative regime, one can define the projectile perturbation parameter $K_p$ [11] as follows

$$K_p = \frac{Z_t}{Z_p} \frac{v_e}{v_p} \quad (3)$$

Where $Z_t$ and $Z_p$ are the target and projectile atomic numbers, respectively, $v_e$ the mean orbital velocity of the active electron of the projectile ion, and $v_p$ the projectile velocity. Therefore, the $K_p$ value dictates the dynamic condition for a given collision system. In view of this, we have discussed all the results in terms of $K_p$ parameter as given in Table 2 while comparing the theoretical predictions with the experiments.

Table 2: It gives the $K_p$ parameters for all the collision systems used in this work. Here, the $K_p$ values are provided for the lowest and the highest beam energies only as used for a system in an experiment. Here, $n$ the principal quantum number of the electronic shell corresponding to the charge state of the projectile at a certain beam energy, is used to have the orbital velocity of the effective electron.

| $Z_t$ | $Z_p$ | $M_p$ | E (MeV) | $v_p$ (m/s) | n | $v_e$ (m/s) | $K_p$ | remark |
|---|---|---|---|---|---|---|---|---|
| 6 | 6 | 12 | 6 | 9.80E+06 | 2 | 6.54E+06 | 0.67 | Fig.1 |
| 6 | 6 | 12 | 60 | 3.10E+07 | 1 | 1.31E+07 | 0.42 | Fig.1 |
| 6 | 8 | 16 | 10 | 1.10E+07 | 2 | 8.72E+06 | 0.60 | Fig.1 |
| 6 | 8 | 16 | 70 | 2.90E+07 | 1 | 1.74E+07 | 0.45 | Fig.1 |
| 6 | 9 | 19 | 30 | 1.74E+07 | 2 | 9.81E+06 | 0.38 | Fig.1 |
| 6 | 9 | 19 | 120 | 3.48E+07 | 1 | 1.96E+07 | 0.38 | Fig.1 |
| 6 | 14 | 28 | 30 | 1.43E+07 | 2 | 1.53E+07 | 0.46 | Fig.1 |
| 6 | 14 | 28 | 110 | 2.75E+07 | 1 | 3.05E+07 | 0.48 | Fig.1 |
| 6 | 15 | 31 | 72 | 2.11E+07 | 2 | 1.64E+07 | 0.31 | Fig.1 |
| 6 | 15 | 31 | 120 | 2.73E+07 | 1 | 3.27E+07 | 0.48 | Fig.1 |
| 6 | 16 | 32 | 30 | 1.34E+07 | 2 | 1.74E+07 | 0.49 | Fig.1 |
| 6 | 16 | 32 | 150 | 3.00E+07 | 1 | 3.49E+07 | 0.44 | Fig.1 |
| 6 | 17 | 35 | 20 | 1.05E+07 | 2 | 1.85E+07 | 0.62 | Fig.1 |
| 6 | 17 | 35 | 160 | 2.96E+07 | 1 | 3.71E+07 | 0.44 | Fig.1 |
| 6 | 18 | 40 | 30 | 1.20E+07 | 2 | 1.96E+07 | 0.54 | Fig.1 |
| 6 | 18 | 40 | 400 | 4.38E+07 | 1 | 3.92E+07 | 0.30 | Fig.1 |
| 6 | 29 | 63 | 42 | 1.13E+07 | 3 | 2.11E+07 | 0.39 | Fig.1 |
| 6 | 29 | 63 | 150 | 2.14E+07 | 1 | 6.32E+07 | 0.61 | Fig.1 |
| 4 | 14 | 28 | 30 | 1.43E+07 | 2 | 1.53E+07 | 0.30 | Fig.2,6 |
| 4 | 14 | 28 | 110 | 2.75E+07 | 1 | 3.05E+07 | 0.32 | Fig.2,6 |
| 6 | 14 | 28 | 30 | 1.43E+07 | 2 | 1.53E+07 | 0.46 | Fig.2 |
| 6 | 14 | 28 | 110 | 2.75E+07 | 1 | 3.05E+07 | 0.48 | Fig.2 |
| 13 | 14 | 28 | 30 | 1.43E+07 | 2 | 1.53E+07 | 0.99 | Fig.2 |
| 13 | 14 | 28 | 60 | 2.03E+07 | 2 | 1.53E+07 | 0.70 | Fig.2 |
| 13 | 14 | 28 | 110 | 2.75E+07 | 1 | 3.05E+07 | 1.03 | Fig.2 |
| 28 | 14 | 28 | 30 | 1.43E+07 | 2 | 1.53E+07 | 2.13 | Fig.2 |
| 28 | 14 | 28 | 110 | 2.75E+07 | 1 | 3.05E+07 | 2.22 | Fig.2 |
| 47 | 14 | 28 | 30 | 1.43E+07 | 2 | 1.53E+07 | 3.57 | Fig.2 |
| 47 | 14 | 28 | 110 | 2.75E+07 | 1 | 3.05E+07 | 3.73 | Fig.2 |
| 79 | 14 | 28 | 30 | 1.43E+07 | 2 | 1.53E+07 | 6.00 | Fig.2,6 |
| 79 | 14 | 28 | 110 | 2.75E+07 | 1 | 3.05E+07 | 6.27 | Fig.2,6 |
| 6 | 16 | 32 | 64 | 1.96E+07 | 1 | 3.49E+07 | 0.67 | Fig.3,4 |
| 6 | 16 | 32 | 64 | 1.96E+07 | 2 | 1.74E+07 | 0.33 | Fig.3,4 |
| 6 | 17 | 35 | 130 | 2.67E+07 | 1 | 3.71E+07 | 0.49 | Fig.3,4 |
| 6 | 17 | 35 | 130 | 2.67E+07 | 2 | 1.85E+07 | 0.24 | Fig.3,4 |
| 6 | 18 | 40 | 8 | 6.20E+06 | 3 | 1.31E+07 | 0.70 | Fig.5 |
| 6 | 18 | 40 | 16 | 8.77E+06 | 3 | 1.31E+07 | 0.50 | Fig.5 |
| 6 | 18 | 40 | 41.6 | 1.41E+07 | 2 | 1.96E+07 | 0.46 | Fig.5 |
| 6 | 18 | 40 | 165 | 2.81E+07 | 1 | 3.92E+07 | 0.46 | Fig.5 |
| 6 | 29 | 63 | 43 | 1.15E+07 | 3 | 2.11E+07 | 0.38 | Fig.5 |
| 6 | 29 | 63 | 65 | 1.41E+07 | 3 | 2.11E+07 | 0.31 | Fig.5 |
| 6 | 29 | 63 | 100 | 1.75E+07 | 3 | 2.11E+07 | 0.25 | Fig.5 |
| 6 | 29 | 63 | 120 | 1.91E+07 | 2 | 3.16E+07 | 0.34 | Fig.5 |
| 6 | 29 | 63 | 150 | 2.14E+07 | 2 | 3.16E+07 | 0.31 | Fig.5 |
| 6 | 29 | 63 | 65 | 1.41E+07 | 3 | 2.11E+07 | 0.31 | Fig:7 |
| 6 | 36 | 84 | 700 | 4.00E+07 | 1 | 7.85E+07 | 0.33 | Fig.8 |

**2.2 Effect of the target atomic number ($Z_t$) on the equilibrium $\bar{q}$**

Fig. 1 shows a comparison between the experimental data [6] and ETACHA predictions [[10],[11]] for the equilibrium $\bar{q}$ against the beam energy, where the projectiles vary while the target is fixed. Here, the ion beam remains the same with silicon and the targets vary from beryllium to gold. For the low target atomic number 4-6, though the ETACHA23 overestimates the data, but agreement is good between the experimental data and ETACHA4 predictions. As the target atomic number increases, a considerable difference occurs. Trend of both the theories are quite alike. For aluminum, the theories underestimates the data only in the low energy side (< 60 MeV), but they do the same throughout the energy range for nickel and silver data. Whereas, agreement is quite good for the gold data. We examine this picture in terms of $K_p$ parameter to know whether the non-perturbative regime is responsible for the observed differences. Comparing the values of $K_p$ parameters given in Table 2 for different targets, we see that the values are less than 1 for Be, C and Al targets. One expects good agreements between the theory and experiment; however, a departure begins to show for Al target at the low-energy side. The $K_p$ parameter is more than 1 for other targets Ni, Ag and Au and thus the agreement between the theory and experiment may be worse as evident from the figure. However, the departure is much higher in the low energy side albeit the $K_p$ parameters remain nearly same for all the beam-energies for a particular system.

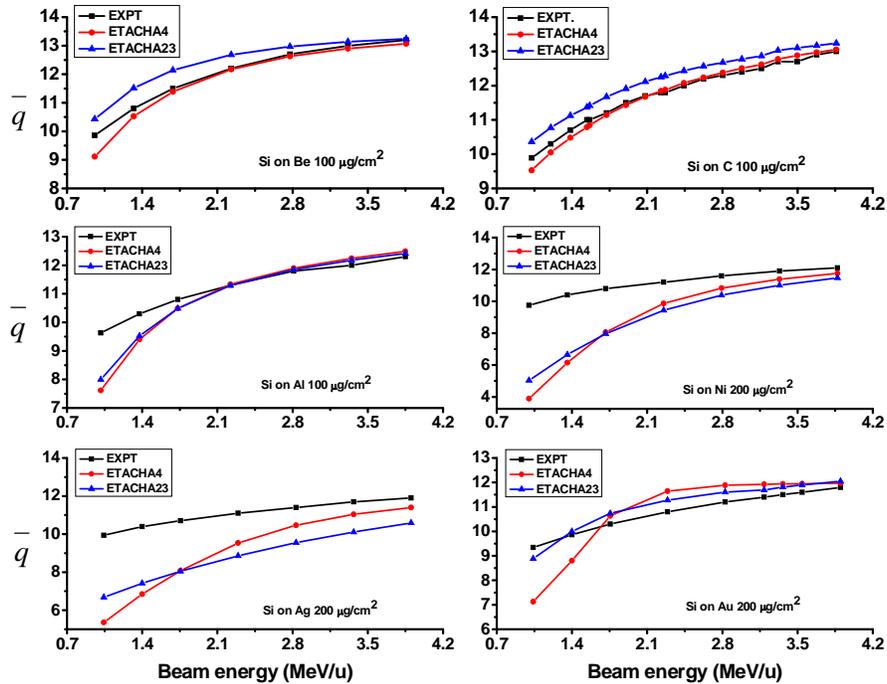

Fig.2: The equilibrium $\bar{q}$ vs beam energy for Si projectiles on different targets. Experimental data are taken from [6]. The solid lines are to guide the eye only.

**2.3 Effect of the projectile charge state on the equilibrium foil thickness**

Fig. 3 shows the effect of the projectile charge states on the equilibrium foil thickness. Experimental data and the ETACHA predictions of the equilibrium $\bar{q}$ have been plotted as a function of the carbon target thickness for 64 MeV sulfur ion with the projectile charge states varying from $6^+$ to $16^+$. Both the experimental data and the theoretical predictions follow the similar trends. Therefore, the ionization and excitation cross sections estimated by the PWBA for ETACHA23, and CDW-EIS and SEIK for ETACHA4 model show similar trend. The equilibrium thickness for projectile charge states with $6^+$ to $14^+$ is smaller (about 10 µg/cm$^2$) than that for the projectile charge states with $15^+$ to $16^+$. The equilibrium thickness for these charge states occurs at much higher thickness ~ 200 µg/cm$^2$. This value is in good accord with the experimental data.

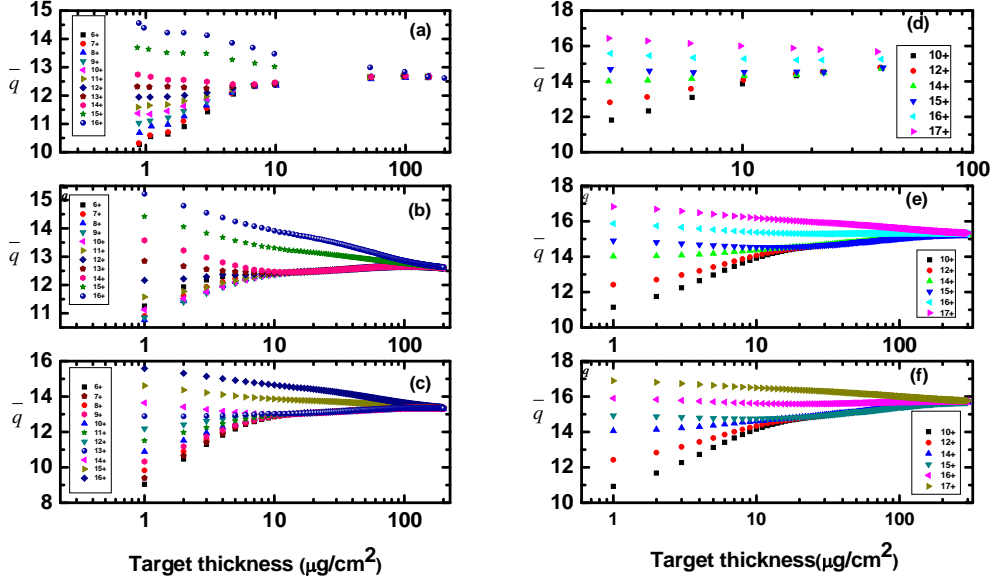

Fig.3: In left panel, the $\bar{q}$ vs carbon target thickness for 64 MeV S ions with $6^+$ - $16^+$. a) the experimental data taken from [17], b) the calculated data from ETACHA4 [11] and c) the calculated data from ETACHA23 [10] and in right panel, the $\bar{q}$ vs carbon target thickness for 130 MeV Cl ion with $10^+$ - $17^+$. d) Experimental data taken from [18], e) ETACHA4 [11] calculation, and f) ETACHA23 [10] calculation.

The above picture is further verified through 130 MeV chlorine ions. Experimental data and the predictions from ETACHA4 and ETACHA23 for the $\bar{q}$ as a function of the carbon target thickness for 130 MeV chlorine ions with the projectile charge state ranging from $10^+$ to $17^+$ have been given also in Fig.3. Both the experimental data and the theoretical predictions follow the similar trends like that for S ions. The equilibrium thickness for the projectile charge states with $10^+$ to $15^+$ is reached at about 20 µg/cm$^2$. The target thickness used in the experiments was only up to 40 µg/cm$^2$, thus it has not achieved the equilibrium thickness for the projectile charge states with $16^+$ to $17^+$. According to the theoretical predictions the equilibrium thickness for the projectile charge states with $16^+$ to $17^+$ may occur at about 300 µg/cm$^2$.

## 2.4 Role of shell effects on the equilibrium thickness

Fig.4 shows the dependence of incident charge states of 64 MeV sulfur [17] and 130 MeV chlorine ions [18]. The $\bar{q}$ values for carbon foil thickness of 10 µg/cm$^2$ (non-equilibrium) for the projectile charge states $6^+$-$14^+$ for sulfur ions and $10^+$-$15^+$ for chlorine ions on 40 µg/cm$^2$ (non-equilibrium) show the effect of shell structure of the projectile ions. At the non-equilibrium thickness, $\bar{q}$ for $15^+$ and $16^+$ in the case of S ion, and $16^+$ and $17^+$ in the case of Cl ions, increase rapidly. This picture reflects the fact that the thickness used are equilibrium thickness for lower charge states, but equilibrium thickness for the two highest charge states is much higher. When the equilibrium thickness required for the highest projectile charge states is used in the experiments, the effect of shell structure disappears fully. The $K_p$ parameters for the projectile with the charge states $6^+$-$16^+$ for sulfur ions and $10^+$-$17^+$ for chlorine ions are all less than 1, but the value for the charge states $15^+$-$16^+$ for sulfur ions (0.67) is about double of that for the charge states $6^+$-$14^+$ (0.33). The case is same for chlorine ions too. Hence, the increase in $K_p$ values can be responsible for the shell effects. The difference in equilibrium thickness seen in Fig. 3 are nothing but the manifestation of the shell effects. Though the general trend in the two versions of ETACHA is seen to be similar, but difference exists on the absolute value of $\bar{q}$ between the predictions of the two codes. ETACHA23 overestimates the data, whereas ETACHA4 represent the data quite well.

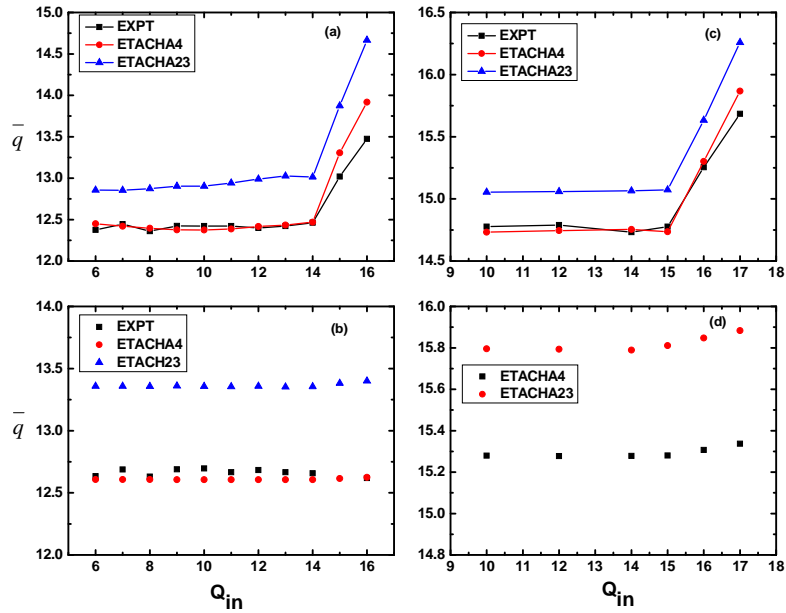

Fig.4: Dependence of the $\bar{q}$ on the incident charge state of the projectile ions; a) for 64 MeV S ions on 10 µg/cm², b) for 64 MeV S ions on 200 µg/cm², c) for 130 MeV Cl ions on 40 µg/cm² and d) for 130 MeV Cl ions on 300 µg/cm². The solid lines are to guide the eye only.

## 2.5 Comparison of the charge state fractions between the experiments and ETACHA predictions

We have noticed in Fig.1 that the experimental $\bar{q}$ differs considerably as the atomic number of the projectile ions increases. Here we elaborate the comparison with two examples through argon and copper beams. Fig. 5 compares the experimental charge state fractions with the ETACHA predictions at four different beam energies of the argon ions and at five different beam energies of the copper ions on carbon target of equilibrium thicknesses. Fig. 5 clearly shows that ETACHA23 overestimates the data throughout the range. Whereas the ETACHA4 prediction underestimates the experimental charge state distribution curves at low beam energies. However, as the beam energy increases the agreement improves gradually, for example, the agreement is good for 165 MeV argon ions. However, no experimental data is available for copper ions beyond 150 MeV energies to know exactly where the agreement may occur.

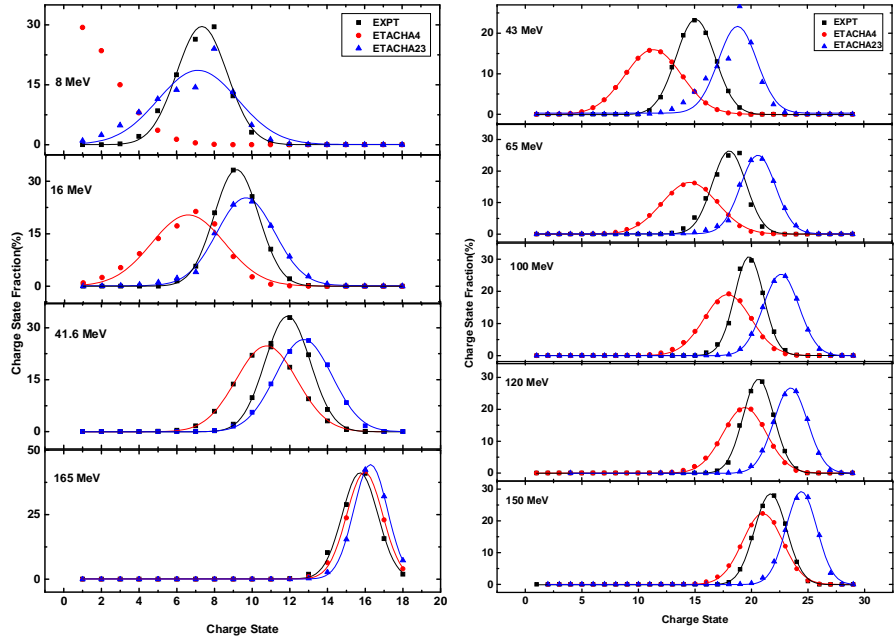

Fig.5: Comparison of charge state fraction between experimental data [19] and ETACHA [[10],[11]] prediction at different beam energies: left panel is for Ar projectiles and right panel is for Cu projectiles on carbon target at equilibrium thickness. Both the experimental and theoretical data points are fitted with the Gaussian functions.

## 2.6 Predictability of the equilibrium target thickness by ETACHA

Though we have seen very good agreement between the theory and experiment for the equilibrium target thickness in case of sulfur and chlorine beams on the carbon target, we make a thorough test on the predicting power of ETACHA codes for same, but this time various target elements are used at different beam energies, for example, a light element beryllium and a heavy element gold. The silicon ion beams in the energy range of 25-110 MeV are used for availability of the data [6] as given also in Fig. 2. The ETACHA predictions for the beryllium and gold target are shown in Fig. 6(a-d). The prediction of ETACHA4 shown in Fig. 6(a) and that of ETACHA23 shown in Fig. 6(b) for beryllium target give clear indication that the equilibrium target thickness increases with increasing beam energy. In contrast, an unusual picture is noticed with the gold target. In this case, the ions at low energies up to 64 MeV do not reach to the equilibrium thickness within 200 µg/cm$^2$, while the equilibrium thickness for ions at higher energies from 79 MeV onwards is achieved about 125 µg/cm$^2$ for ETCHA4 and 175 µg/cm$^2$ for ETCHA23. Further, the experimental $\bar{q}$ data [6] are in good agreement with the ETACHA predictions only for the beam energies ≥ 79 MeV. The incident charge state varies with the beam energy and it lies in the range of 5-10 for the beam energies 26.8 to 108 MeV and hence the shell effect does not play any roles over here. Further, both the codes give similar results, hence, different pictures obtained in beryllium and gold cannot be attributed to the difference of the $K_p$ values as well; see Table 2.

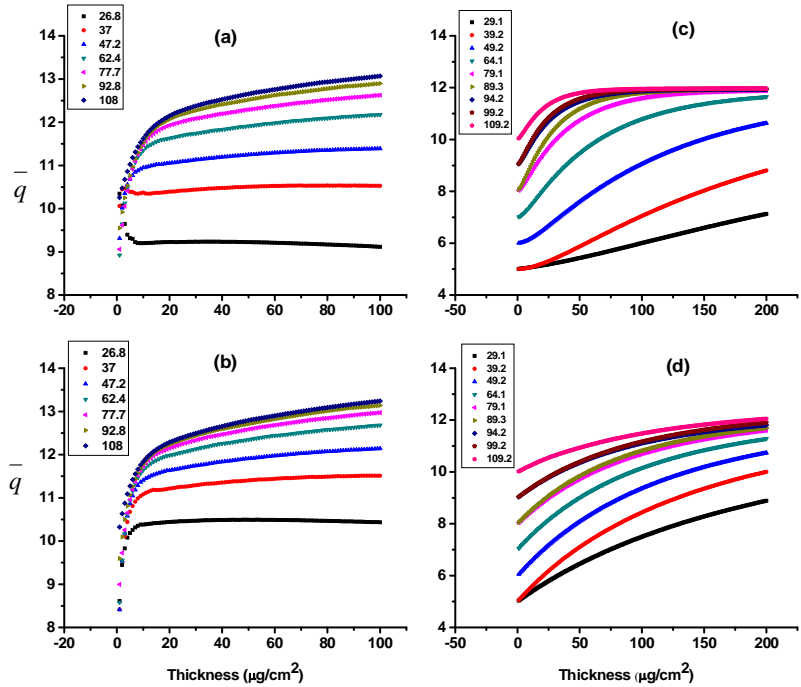

Fig.6: Theoretical $\bar{q}$ vs target thickness for different silicon ion beam at different energies, a) ETACHA4 [11] predictions, b) ETACHA23 [10] predictions for beryllium target, c) ETACHA4 [11] predictions and d) ETACHA23 [10] predictions for gold target.

In Fig.7, we have plotted $\bar{q}$ against the carbon target foil thickness for 65 MeV Cu$^{+9}$ ion beam. Experimental $\bar{q}$ data are compared with ETACHA4 and ETACHA23 predictions in Fig.7(a) and 7(b), respectively. Here, we notice that neither ETACHA4 nor ETACHA23 reproduces the experimental data, where the ETACHA4 underestimates and the ETACHA23 overestimates the experimental data. Note that the ETACHA23 predictions represents the similar trend with the experimental data. The perturbation parameter for this collision system is only 0.31, hence the cross sections obtained by plane wave Born approximation used in ETACHA23 and CDW-EIS and SEIK models used in ETACHA4 should not differ much from each other. However, the substantial difference between the two model predictions raises a question difficult to answer at this stage.

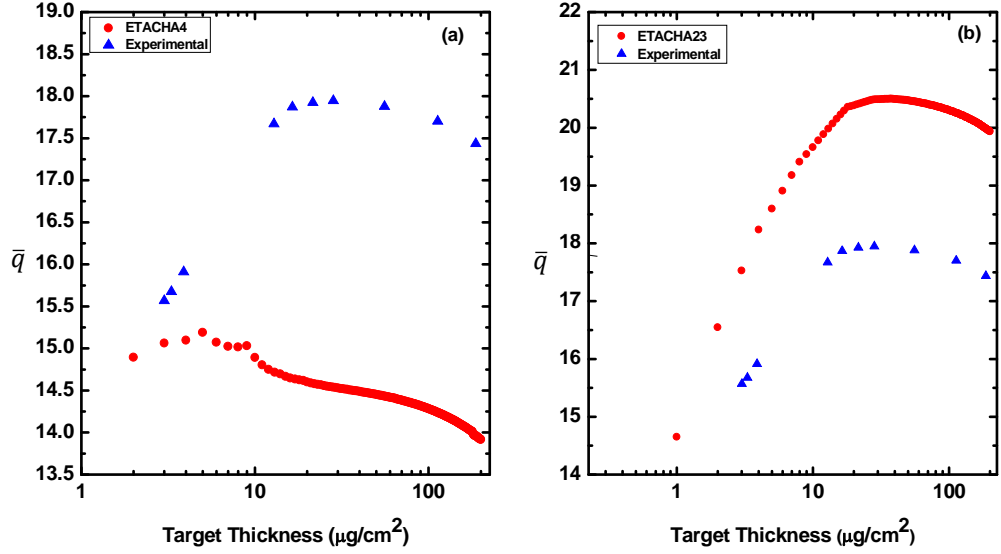

Fig.7 : Experimental $\bar{q}$ data [20] vs carbon target thickness for 65 MeV $Cu^{+9}$ ion beam are compared with (a) ETACHA4 and (b) ETACHA23 predictions.

Another case of 700 MeV $Kr^{+34}$ on carbon target is taken to study the charge equilibration as shown in Fig.8 (a-c), where Fig.8(a) for the experimental results, Fig.8(b) for the ETACHA4 and Fig.8(c) for ETACHA23 predictions. One can see clearly that the equilibration is reached in the experiment about 200 µg/cm$^2$, whereas the same is reached at almost double the thickness in the ETACHA4 calculation. In contrast, the equilibration arrives at smaller thickness about 50 µg/cm$^2$ in the ETACHA23 calculations, but the $\bar{q}$ keeps on increasing with the thickness. Both the experimental and theoretical curves can be fitted with the following equation as given by Kanter et al [21].

$$\bar{q} = q_\infty + (q_0 - q_\infty) \exp\left(\frac{-x}{\lambda}\right) \qquad (4)$$

Where $q_\infty$ is the equilibrium charge state, $q_0$ is the incident charge state ($34^+$) and $\lambda$ is the mean free path. The measured and theoretical $q_\infty$ values are quite close, but the theoretical mean free path with the ETACHA4 is 2.89 times the experimental mean free path. This implies that the theoretical total cross section is 0.35 times the measured total cross sections. Hence, the total cross sections estimated from the SEIK and CDW-EIS approximations are under estimated. Whereas the theoretical mean free path with the ETACHA23 is 0.73 times the experimental mean free path or the theoretical total cross section is 1.36 times the measured total cross sections.

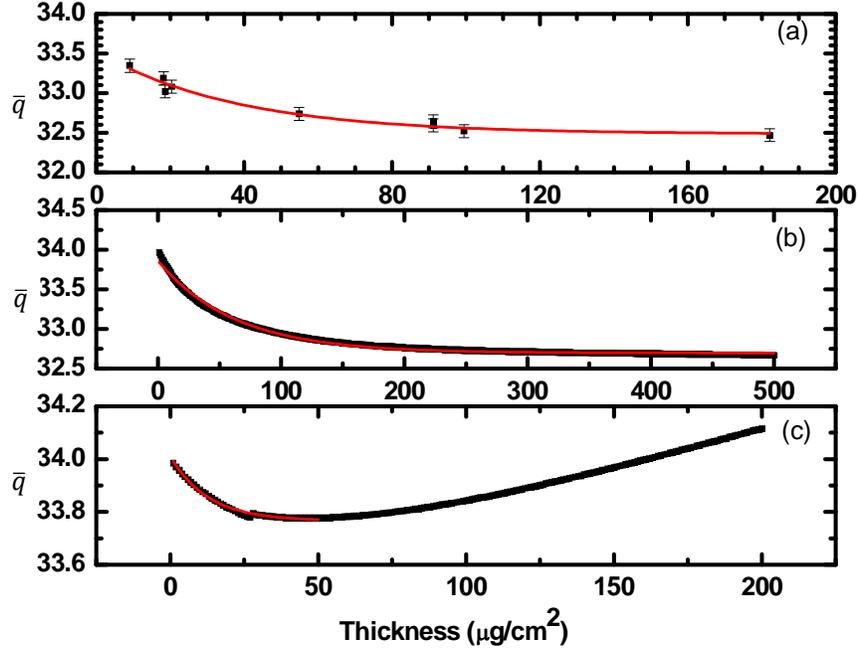

Fig.8: The $\bar{q}$ vs carbon target thickness plot for 700 MeV $Kr^{+34}$ ion beam: (a) Experimental data taken from [21] and (b) ETACHA4 predictions. Both the experimental and theoretical data points are fitted with the equation (3) and shown in solid red lines. Note that for fitting the ETACHA23 data we have considered the data up to 50 µg/cm².

2.7. **Effect of the electron capture at the surface**

Apart from the electron capture processes in the bulk of the foil, certain electron capture processes occur at the exit surface of the foil also [[22],[23],[24],[25]], including the radiative as well as non-radiative capture processes. The radiative capture leads to radiative electron capture (REC), which is included in the ETACHA calculations. The non-radiative capture results in population of the Rydberg states, which are yet to be considered in the theory. Whatsoever, the surface electron capture processes must alter every charge state produced in the bulk. Consequently, the charge distribution may shift towards the lower charge states. The effect can be more prominent in low energy side as the electron capture cross section is high at the low beam velocities [22]. Exactly this picture i.e., the predictions are higher than the measured values, is being reflected in the ETACHA23 data shown in Fig. 1 & 5, because the theories have not taken yet the surface effects into account. Even though the electron capture processes at the exit surface are not taken also into consideration in the ETACHA4, still in many cases (see Fig. 1 & 5) the predicted values are less than the measured data.

Only difference between the theoretical approaches in ETACHA23 and ETACHA4 is that the former makes use of the PWBA for accounting the ionization and excitation processes; whereas the later applies the CDW-EIS approximation [[12],[13]] was applied for obtaining the ionization cross sections and the SEIK model [[14],[15]] for the excitation cross sections. Hence, CDW-EIS and SEIK may underestimate the ionization and excitation cross section, respectively.

The difference in $\bar{q}$ between the experiment and ETACHA23 shown in Fig. 1-4 & 6-8 goes even up to 4. Such scenario favors the multi-electron capture process, which is observed in low to high energies in several cases [[26],[27],[28]]. Estimating the multi electron capture processes is beyond the scope of this work. However, we take an attempt to understand the trend considering only the single-electron capture processes. Schlachter et al [29], formulated a universal scaling rule for estimating the single electron-capture cross sections for fast (0.3 to 8.5 MeV/u), highly charged ions in atomic and molecular targets. This rule predicts well the capture cross sections for a wide variety of projectile-target combinations. Bulk of the foil consists of many atomic layers, whereas the surface only a few layers. Hence, the ion-solid (bulk) interactions are multi-collisional phenomenon and the ion-surface collisions can be treated as single collisional events. Hence, the rule built for atom or molecular target can be used to represent the ion-surface interactions. Thus, we have made use of the capture cross section formula [29] in case of the non-radiative capture processes at the exit surface by the projectile ions as follows

$$\sigma = 1.1 \times 10^{-8} q^{3.9} Z_2^{4.2}/E^{4.8} \quad \text{cm}^2 \qquad (5)$$

Where $q$ is incident charge state of the projectile; in the present case the incident charge state is the mean charge state produced in the bulk as predicted by ETACHA23 or ETACHA4, $Z_2$ is atomic number of the target atom and E is the beam energy in keV/u.

The capture cross section obtained from the equation (5) have been plotted for several systems as used in Fig. 1 and shown in Fig. 9. Fig. 1 shows that the measured $\bar{q}$ differs considerably from the ETACHA23 predictions and the differences increase at the low beam energies. Similar trend is also found in Fig. 9 that the electron capture cross section increases by a few orders of magnitude at the low beam energies. Higher $\bar{q}$ in ETACHA23 is bound to be lowered, once the surface electron capture is incorporated in the calculations. In next step, the systems used for Fig. 2 are used to have the capture cross section as a function of the beam energies and shown in Fig. 10. Though the surface electron capture phenomenon can explain the fact observed with Si-ions colliding with Be and C targets, the rest shows more complex nature because of the nonperturbative processes playing their vital role along with the surface electron capture.

Electrospray ionization (ESI) of big molecules like protein produces a broad distribution of multiply charged ions [30], however the understanding of the exact mechanism is far from reach [31]. We believe the surface potential can play important role in electrospray ionization (ESI) in the following way. When a charged molecule approaches the capillary surface, an image potential is generated. The potential can drag the ion and thus enforce the ion to lose certain kinetic energy. It means the ion undergoes certain ionization and thus forms multiply charged ions in the ESI until the Raleigh limit is attained. A fact is noteworthy that the ion gets ionized by the potential created by itself called a self wake [32]. Hence, the self wake induced ionization is responsible in creating high charged states in the ESI.

It is clear from the above scenario that the electromagnetic measurements, magnetic or electric, cannot map the actual picture occurring in the bulk of the foil as it concerns with the total charge of the ion. The total charge is governed by both the bulk and surface effects. Another technique involving the x-rays [33] are used to study the charge state distributions in space [34] and laboratories [35]. The prompt x-ray technique using the characteristic x-ray lines avoids the long-lived Rydberg states generated from exit surface [[25],[36]] and thus is appropriate to study the charge state distribution only in the bulk of the foil. In order to take the charge changing processes in the bulk as well as REC processes at the exit surface, the measurement using the REC peak structures can be considered [36].

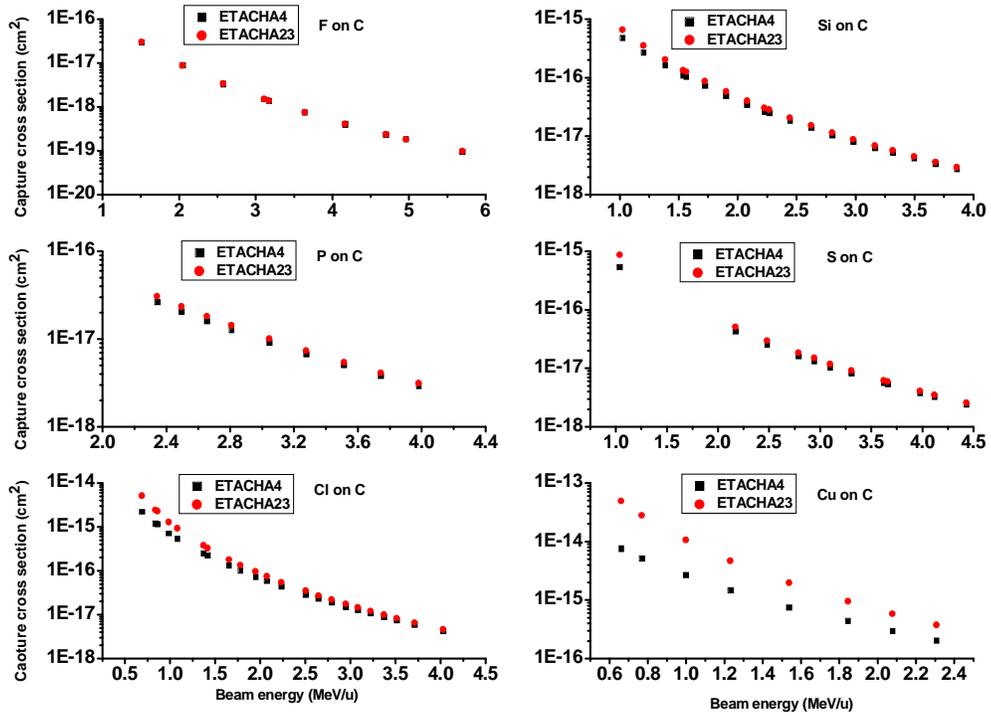

Fig. 9 : The electron capture cross section obtained from equation (5) vs beam energy for different projectile ions as used for Fig. 1 on carbon target. Red and black symbols indicate whether the incident charge state used in equation (5) is taken from ETACHA23 or ETACHA4.

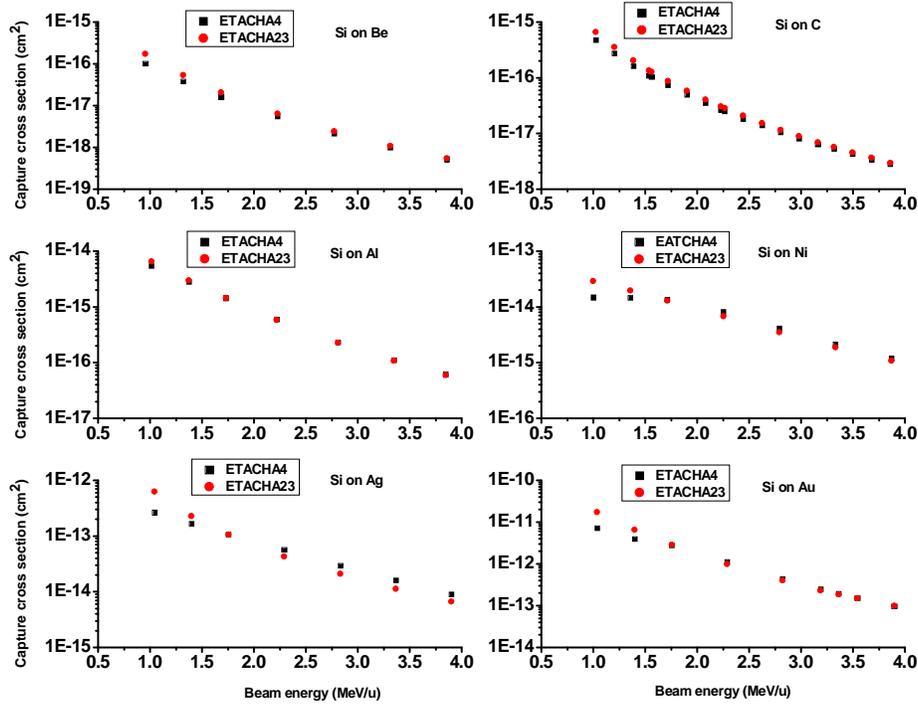

Fig. 10 : The electron capture cross section obtained from equation (5) vs beam energy for Si ions on different targets as used for Fig. 2. Red and black symbols indicate whether the incident charge state used in equation (5) is taken from ETACHA23 or ETACHA4.

### 3.Conclusions:

The range of validity of the ETACHA codes have been examined in the low energies (E < 10 MeV/u) for the projectiles with a few electrons to 20 electrons. We have illustrated the merits and demerits of the two different versions of the code by comparing the computed data with the available experiments as given in Table 3. In particular, we see that ETACHA23 overestimates the $\bar{q}$ data, whereas ETACHA4 reproduces the experimental data quite well for $Z_p = 10 - 16$, however somewhat small departure appears for $Z_p = 6 - 9$ and large departure (predictions being smaller) becomes visible from $Z_p \geq 17$ even though the perturbation parameters lie in the range of 0.3 - 0.7 (Fig.1). In order to further study on the large departure for heavier projectiles, we have studied it through the charge state distribution for a wide range of $K_p$ parameters. It is noticed that the charge state distribution curve reproduces well the experimental fact if $K_p \leq 0.3$ (Fig. 5). However, an acute problem occurs while the $\bar{q}$ is studied through 65 MeV $Cu^{+9}$ and 700 MeV $Kr^{+34}$ as a function of the carbon foil thickness. Even though the $K_p$ parameter ~ 0.3, neither ETACHA23 nor ETACHA4 can represent the data of these experiments; the former model overestimates while the later underestimates the measured data (Fig. 7). Further, the equilibrium thickness of ETACHA4 predictions is about twice of the measured value. The reason of such difference may be attributed to the fact that the total collision cross section used in ETACHA4 is 0.35 times the measured value. In contrast, the total collision cross section used in ETACHA23 is found to be 1.36 times the measured value.

Table 3: Comparison of the experimental results with ETACHA23 and ETACHA4 are summarized. Observed agreements and disagreements are discussed in the light of nonradiative capture (NRC) processes at the exit surface of the target foil.

| Physics issue | ETACHA23 | ETACHA4 | Remarks |
|---|---|---|---|
| Effect of the projectile atomic number ($Z_p$) on the equilibrium $\bar{q}$ in reference to the results shown in Fig.1. The perturbation parameters lie in the range of 0.3 - 0.7 for the collisions systems used. | 1) Calculated $\bar{q}$ data overestimate the experimental data for projectile ions C, O and F. <br> 2) Calculated $\bar{q}$ data overestimate the experimental data for projectile ions Si, P and S. <br> 3) Calculated $\bar{q}$ data overestimate the experimental data for projectile ions for Cl, Ar and Cu ions. | 1) Here also calculated $\bar{q}$ data overestimate the experimental data for projectile ions C, O and F. <br> 2) Calculated $\bar{q}$ data are in good accord with the experimental data for projectile ions Si, P and S. <br> 3) Calculated $\bar{q}$ data underestimate the experimental data for projectile ions Cl, Ar and Cu ions. | ETACHA23 overestimates the $\bar{q}$ data, whereas ETACHA4 either overestimates or shows good agreement or even underestimates the data. Inclusion of the NRC process in the theory will reduce the calculated $\bar{q}$ values; thus higher values will tend to match with the measured values. On the other hand, equal or lower values at present will move farther to the lower values. Hence, ultimately ETACHA23 will represent the data better. |
| Effect of the target atomic number ($Z_t$) on the equilibrium $\bar{q}$ for Si ions in reference to the results shown in Fig.2. | 1) Calculated $\bar{q}$ data overestimate the experimental data for low atomic number target $Z_t = 4 - 6$. ($K_p$ parameter 0.3-0.48) <br> 2) Calculated $\bar{q}$ data agree well with the experimental data beyond 2.3 MeV/u, but underestimate at low energies for Al target. ($K_p$ parameter 0.7-1.03) <br> 3) For Ni and Ag targets, theoretical predictions underestimate the experimental data throughout the energy range. ($K_p$ parameter 2.13-3.73) <br> 4) For Au target, predictions slightly overestimate the experimental data throughout the energy range. ($K_p$ parameter 6.0-6.27) | 1) For low atomic number target $Z_t = 4 - 6$, the agreement between experiment and prediction is good. <br> 2) For Al targets, the predictions are the same as that of ETACHA23. <br> 3) For Ni and Ag targets, the predictions underestimate the experimental data like the ETACHA23. <br> 4) For Au target, the predictions overestimate the experimental results for energy $E > 1.7$ MeV and underestimate at the low energies. | With respect to the discussion above that overestimation on $\bar{q}$ is quite obvious and can be taken as good results, however, under estimation is a failure. Hence, ETACHA23 explains well the Si ion on target with $Z_t = 4 - 6$, but fails to reproduce the experiments for $Z_t \geq 13$, where the collision dynamics is in the nonperturbative region. While ETACHA4 fails even for $Z_t = 4 - 6$. |
| Effect of the projectile charge state on the | The predictions follow the experimental data very well. | The predictions are exactly same as ETACHA23. | We have seen above that absolute value of calculated $\bar{q}$ is different from the |
| equilibrium foil thickness in reference to the results shown in Fig.3. ($K_p$ parameter 0.24-0.67) | | | measured values, nevertheless absolute value of $\bar{q}$ does not make any role in finding the equilibrium foil thickness. Hence, both ETACHA23 and ETACHA4 predicts the experimental data for S and Cl ion on C-foils. |
| Role of shell effects on the equilibrium thickness in reference to the results shown in Fig.4. ($K_p$ parameter 0.24-0.67) | It predicts the shell effect very well, however, calculated $\bar{q}$ data as a function of $q_{in}$ overestimate the experimental data for both S and Cl ions. | It also predicts the shell effect very well and the predictions on $\bar{q}$ represent the experimental data well for both S and Cl ions. | Both the versions explain well the shell effects. As explained above that the overestimation on $\bar{q}$ is good and even agreement is not so. Hence, inclusion of NRC will make ETACHA23 data close to the experiments and ETCHA4 data will move far off. It implies the ionization cross sections underestimated. |
| Comparison of the charge state distributions for Ar and Cu ions at different beam energies ($K_p$ parameter ~ 0.3) in reference to the results shown in Fig.5. | For both Ar and Cu ions, the predictions overestimate the experimental data throughout the energy range, however, the difference reduces with the beam energy. | For both Ar and Cu ions, the predictions underestimate the experimental data, however, the difference reduces with the beam energy. | Here also, inclusion of NRC will make ETACHA23 data close to the experiments and ETCHA4 data will be farther off towards lower side. It gives clear indication that CDW-EIS underestimates the ionization cross section to a great extent; hence frozen core approximation may not be suitable. |
| Obtaining the equilibrium target thickness from $\bar{q}$ versus target thickness plot as the results shown in Fig.6-8. | 1) The predictions show for Si ions on Be target that equilibrium thickness increases with the beam energy as expected (Fig.6, $K_p$ parameter 0.3-0.32). <br> 2) In case of Au target, equilibrium thickness cannot be reached within 200 µg/cm² for beam energies below 79.1 MeV and the equilibrium thickness is about 175 µg/cm² for higher beam energies (Fig.6, $K_p$ parameter 6.0-6.27)). <br> 3) For Cu ions on C target, though it predicts well the equilibrium thickness, but | 1) Here also the predictions are same as ETACHA23 (Fig.6). <br><br> 2) The scenario is same as ETACHA23 for beam energies below 79.1 MeV and the equilibrium thickness is about 125 µg/cm² for higher beam energies (Fig.6). <br> 3) For Cu ions on C target, predictions show a awkward trend (Fig. 7). <br> 4) In this case, the equilibrium | Both the versions fail to predict the equilibrium thickness for the case of Si ions on Au target (Fig.6). The $K_p$ parameter for the Cu ion beam at 65 MeV on C target is 0.31, still ETACHA4 underestimates by 0.5 to 3.5 unit of charge states with increasing foil thickness. Frozen core approximation in CDW-EIS and SEIK may be responsible. In contrast, ETACHA23 explains the equilibrium thickness very well as electron capture process at the exit surface ought to reduce the |
| | overestimates the $\bar{q}$ data (Fig. 7, $K_p$ parameter 0.31).<br>4) For Kr ions on C target, the experimental equilibrium thickness is about 200 µg/cm², whereas predictions result at about 50 µg/cm². Hence, it underestimates the measurements by four times (Fig.8, $K_p$ parameter 0.33). Calculated $\bar{q}$ data beyond 50 µg/cm² show a upward trend, which is very unlikely. | thickness is reached at about 400 µg/cm².Thus, it overestimates the measurements by two times (Fig.8). | calculated $\bar{q}$ data, but will not affect the equilibrium thickness. Inclusion of NRC at the exit surface will bring the predictions close to experiments (Fig.7). Frozen core approximation does not pose any problem for high beam energy of $Kr^{34+}$ (700 MeV) on C target and thus data are well represented by ETACHA4. Further, inclusion of NRC at exit surface will make the agreement even better. Though ETACHA23 predictions do not follow well the data, but NRC at exit surface will improve to a good extent (Fig.8). |
| Effect of the electron capture at the surface | This effect is not considered in ETACHA23. | This effect is not considered even in recent ETACHA4 versions also. | Inclusion of electron capture at the exit surface will certainly take the calculated $\bar{q}$ and CSD data towards lower side but will not alter the equilibrium thickness predictions. |

Since the silicon projectiles show good agreement on the carbon target (Fig.1), we have made a thorough check on various other targets ranging from $Z_t = 4 - 79$. The departure seen at the low energies (Fig. 2) cannot be explained alone with the $K_p$ parameters as though the $K_p$ value is somewhat high for high $Z_t$ but do not vary much with the energies (Table 2). In order to elucidate this fact, the electron capture at the exit surface has been considered. Though REC is included in the ETACHA calculations, the non-radiative capture processes leading to the population of Rydberg states have not been considered yet, which again will lower the charge states further. Note that ETACHA4 predictions already are lower than the measured values for $Z \geq 17$. The non-radiative electron capture processes at the exit surface will further lower the values and thereby, the differences will enhance more. Hence, ETACHA4 predictions at the low beam energies remain questionable. In contrast, the ETACHA23 predictions are rather higher than the measured values and inclusion of non-radiative electron capture cross sections at the surface as shown in Fig. 9 will bring the values closer to the measured data.

Both ETACHA23 and ETACHA4 reproduce well the dependence of $\bar{q}$ on the projectile charge states and target foil thickness up to $Z_p \sim 17$ (Fig.3&4). Especially, these models imitate the experimental fact that the equilibrium thickness does vary on the projectile charge states because of the shell effects of the projectile ions (Fig 3 &4). However, the theories fail to represent the $\bar{q}$ data for heavy projectiles for example 65 MeV Cu on carbon target despite the $K_p$ parameter ~ 0.3 i.e., belonging to the perturbative regime. The difference is about 4 units of charge (Fig.7) between ETACHA23 and the measured data. Hence, nonradiative multi electron capture taking place at the exit surface in the influence of wake and dynamic screening effects [37]. This can be a possible mechanism of multiply charge formation in the electrospray ionization of big molecules.

**Acknowledgements:** It is also our pleasure to thank Prashant Sharma for illuminating and frequent discussions during the course of this work.